\documentclass[12pt,a4]{article}
\usepackage{amssymb,amsmath} 
\usepackage[dvips]{graphicx}

\begin{document}

\newcommand{\fr}[2]{\frac{{\displaystyle #1}}{{\displaystyle #2}}}
\newcounter{enumct}
\newenvironment{Enumerate}{\begin{list}{\arabic{enumct}.}%
{\usecounter{enumct}\setlength{\topsep}{0.2mm}%
\setlength{\partopsep}{0.2mm}\setlength{\itemsep}{0.2mm}%
\setlength{\parsep}{0.2mm}}}{\end{list}}


\begin{center}
{\Large \bf

\boldmath The study of triple gauge boson anomalous interactions via
process $\mathbf{\lowercase{e}^-\gamma\to W^-\nu}$. Leptonic
$W$ decay mode. } \\

\vspace{5mm}
 To be published at PHOTON'2001 and QFTHEP proceedings.

\vspace{4mm}


Dmitriy A. Anipko, Ilya F. Ginzburg, Alexey V. Pak\\

\vspace{4mm}

Sobolev Institute of mathematics SB RAS,  prosp. Koptyuga 4\\
and Novosibirsk State University, ul. Pirogova 2\\
Novosibirsk, 630090, Russia\\
E-mail: ginzburg@math.nsc.ru

\end{center}


\begin{abstract}

We study possibilities to measure the triple anomalous W-boson
couplings to photon in  the $e\gamma\to W\nu$ process via its
lepton decay channel (with the simplest signature). We found
that in the study of the $W$ quadruple momentum $\lambda$ one can
limit himself within a small region in phase space. A way to find
this region is proposed. The obtained estimates for $\lambda$ at
TESLA project are roughly twice better than anticipated for
$e^+e^-$ mode. For $W$ anomalous magnetic momentum the discussed mode
gives no improvements as compared $e^+e^-$ mode.

\end{abstract}

Study of anomalous interactions of gauge bosons (beyond the
Standard Model -- SM) is an essential part of the program of
Linear Colliders (LC), both in $e^+e^-$ and $e\gamma$, $\gamma\gamma$
modes (Photon Colliders)
\cite{GKST},\cite{TESLATDRVI}. The $e^+e^-$ mode of the LC has
been studied thoroughly  \cite{TESLA}. The process $e\gamma\to
W\nu$ was considered in respect to Photon Collider program in 1984
\cite{GKPS} first. Anomalous gauge boson interactions in
$e^+e^- \to W W$, $\gamma\gamma\to W W$, $e\gamma \to W \nu$ processes
have been studied in the papers \cite{yehudai91},\cite{schrempp91}
with neglected backgrounds, W--boson decay and involving initial
particles' spectra, polarizations and luminosities far from modern
understanding.

The $e\gamma\to W\nu$ process has the following advantages as compared
to as compared to the $e^+e^-\to WW$:

(a) much higher cross section not falling down at higher energies,
as compared to the $\sigma(e^+e^-\to WW)$, which decreases with energy;

(b) only $\gamma WW$ anomalies contribute here, making analysis
more definite comparing with $e^+e^-\to WW$ case when $ZWW$
anomalies influence as well.

We use the standard effective lagrangian parameterization with
anomalous parameters  $\Delta k$ and $\lambda$ -- anomalous
magnetic and quadruple momenta of $W$--boson respectively as
$$
    e[{W^\dagger}_{\mu\nu} W^\mu F^\nu -
        {W^\dagger}_\mu F_\nu W^{\mu\nu} + 
        (1+\Delta k) {W^\dagger}_\mu W_\nu F^{\mu\nu} +
        \frac{\lambda}{m^2_W} {W^\dagger}_{\lambda\mu} {W^\mu}_\nu F^{\nu\lambda}
    ]\,.
$$ \vspace{3mm}

{\bf\boldmath Some important features of the $e\gamma\to W\nu$
reaction} can be seen before numerical simulations:\\
 $\bullet$ Since only left--hand polarized fermions interact in the
$We\nu$ vertex, the cross section is proportional to $(1-2\lambda_e)$
where $\lambda_e$ stands for the degree of electron longitudinal
polarization. Thus, varying the mean electron helicity, one can
measure the right current admixture in this vertex in the new
region of $W$ virtualities.\\
 $\bullet$ In our problem we assume anomalous effects to be
relatively weak. Therefore, in the observable variations of
cross sections only linear by $\Delta k$ and $\lambda$ effects
should be considered to be experimentally observable. The
structure of helicity amplitudes for $e\gamma\to W\nu$ process
(without decay) shows that for the left hand or right hand
polarized initial photons both anomalies ($\Delta k$ and
$\lambda$) contribute to the cross sections while for
unpolarized photons linear on $\lambda$ effects are canceled.
Therefore, the analysis with unpolarized photons is incomplete.

To analyze the process we considered observable channels after
$W$ decays. Note that the description of these channels contains
additional diagrams (not only $e\gamma\to W\nu$ with
subsequent decay). For example, the muon decay channel contains a
diagram in which an initial photon interacts with the muon after
$W$ decay.

We classify {\bf the observable channels of the reaction} by
the
\begin{table}[hbt]
\begin{tabular}{|c|c|c|c|}\hline
\multicolumn{2}{|c|}{muon(electron) channel}&$\tau$ -- channel&
quark channel\\\hline
  1&2&&\\\hline
   $\begin{array}{rl}
        e\gamma\to & W^-\nu_e \\ & \downarrow \\
        & \mu (e) \bar{\nu}_\mu
    \end{array}$ &
    $\begin{array}{rl}
        e\gamma\to & W^- \nu_e \\ & \downarrow \\
         & \tau\bar{\nu}_\tau \\
        & \downarrow \\ &
        \mu (e) \bar{\nu}_\mu\nu_\tau
    \end{array}$&
   $\begin{array}{rl}
        e\gamma\to & W^- \nu_e \\ & \downarrow \\ &
        \tau\bar{\nu}_\tau \\
        & \downarrow \\  & \nu_\tau+hadrons
    \end{array}$ &
    $\begin{array}{rl}
        e\gamma\to & W^-\nu_e \\ & \downarrow \\ & q \bar{q}
    \end{array}$ \\\hline
\end{tabular}
\caption{\it W decay channels.}
\end{table}
observable particle and its origin (Table 1). We distinguish, for
example, two muon channels, where channel 1 corresponds to direct $W$
decay into $\mu\bar{\nu}$, and channel 2 corresponds to cascading
decay to muon and neutrinos with intermediate $\tau$ state.

We consider first only muon channel.

{\bf  Event selection cuts.} We impose two constraints on muon
escape angle $\theta$ and its transverse momentum $p_\bot$ (for
operations with $\sqrt{s} \le 1$ TeV):
 $$
\boxed{\mbox{1. $\pi - \theta_0 \ge \theta \ge
\theta_0=10\mbox{~mrad}$;\hspace{1.5cm} 2. $ p_\bot > p_{\bot 0} =
10$~GeV.}}
 $$
Condition 1 corresponds to the TESLA detector expected angular
limitation. The second cut allows to exclude or suppress many
background processes. We found that reasonable (not excessive)
increase of $\theta_0$ and $p_{\bot 0}$ influences our results
weakly.

{\bf Background processes} are those where either muon is
only particle that can be observed or where other
charged particles and photons cannot be detected due to
their small escape angles. These are:

1. {\it Processes in which all the final particles can
potentially be observed in principle} --- $e\gamma\to
e\mu^+\mu^-$, $e\gamma\to eZ\gamma$ ($Z\to \mu\bar{\mu}$).

2. {\it Processes including neutrinos in the final state} ---
$e\gamma\to e\bar{\tau}\tau$ ($\tau\to \mu$), $e\gamma\to eZZ$
($Z\to\nu\bar{\nu}$, $Z\to\mu\bar{\mu}$), $e\gamma\to \nu WZ$
($Z\to\nu\bar{\nu}$, $W\to\mu\nu$), $e\gamma\to eW^-W^+$ ($W^+\to
\ell^+\nu_\ell$, $W^-\to\mu{\bar \nu}_{\mu}$).

3. {\it Processes caused by deviation of the state
from ideal due to conversion mechanism.} There happen $e^-e^-\to \nu W^- e^-$
collisions with residual electrons in the photon beam
and $\gamma\gamma\to W^-W^+$ process with beamsstrahlung photons. We
consider in this group also the process $e\gamma\to W\nu$ with
photons from multiple electron scattering on laser photons.

Transverse momentum conservation along with the cuts imposed
exclude processes of the first group (it is impossible that with
the energy and angular cuts given only one observed particle has
transverse momentum higher than $10$~GeV).

Of the third group processes,  $e\gamma\to W\nu$ collisions with
low energy photons from multiple electron scattering have been
simulated and analyzed in detail. As for other processes, they
cannot be excluded only by cuts. But it follows from our analysis
that the anomalies considered can be extracted with good efficiency
from the regions of muon momentum plane
$(p_L,p_\bot)$ close to boundaries of the phase
space permissible in the reaction. These regions are either beyond
the reach of some background processes or their estimated cross
sections are relatively small in these regions.
Therefore, at the first analysis most of the backgrounds
should not have been simulated.

{\bf Main parameters}. In our analysis the electron longitudinal
polarization was taken as $2\lambda_e = -0.85$, luminosity value
in $e\gamma$ mode was considered $1/4$ of its value in $e^+e^-$ mode
\cite{TESLATDRVI} $\int\mathcal{L}_{e\gamma}dt = \frac{1}{4}\int\mathcal{L}_{e^+e^-}dt$.
Different signs of photon circular polarization have been accounted.
$W$ parameters were taken from \cite{PDG}.

We assume main parameter for $e\to\gamma$ conversion $x=4.8$
(corresponding to $E_e= 500$ GeV).  Shape of the high energy part
of the photon spectrum depends weakly on details of conversion,
the beam size and laser flash energy. We used here spectra
from papers \cite{GKPS},\cite{spectrumShape} with parameter $\rho = 1$.
On the contrary, shape of the low-energy part does depend strongly
on interaction details and cannot reliably be determined in advance.
In addition, low-energetic photons are almost unpolarized. To
imitate this part of spectrum, we used the low energy part of
backscattered photons' energy spectrum, given in
the conversion point with completely unpolarized photons. It has been
found that these low energy photons don't influence the results
significantly.

To take into account the electron initial state radiation we used
the effective electron spectrum from
refs.~\cite{fadin85},\cite{jadach91}.

{\bf Calculations.} We calculated distribution of final muons
over components of their momentum $\partial^2\sigma/(\partial
p_\|\partial p_\bot)$ in SM and with anomalies, using CompHEP
package \cite{comphep} for symbolic and numerical calculations.
For the 2nd $\mu$-channel distribution over $\tau$ momenta was
calculated using CompHEP (the same as for channel 1), result was
numerically convoluted with distribution (which is simple to
calculate) of muons from $\tau$ decay with $\tau$ decay
branching ratio from \cite{PDG}. This procedure allowed us to
avoid analyzing the multiparticle final phase space (after
integration over neutrino momenta and averaging over muon spin
this distribution becomes independent on the polarization of
intermediate $\tau$. Indeed, this distribution is determined by
two 4-momenta $p_\tau$ and $p_\mu$ and one pseudovector of
$\tau$ spin $s_\tau$, but one cannot combine a scalar value,
including $s_\tau$ and both momenta). This approximation is
acceptable since $\tau$--lepton width is rather small. We found
that final distributions on muon momentum for these two channels
are similar.

Computed distributions in SM and with anomalous
interactions were used to calculate {\bf Statistical
Significance} ($SS$) defined as:\\

\centerline{ $SS=\fr{N_{(SM+anom)}
 - N_{SM}}{\sqrt{N_{SM}}} $}
The quantity $\sqrt{N_{SM}}$ in denominator corresponds to the
situation where relative influence of anomalies on cross section is small.

{\it On the first step} we put values $\lambda=\lambda_{sim}=
0.1$ or $\Delta k=\Delta k_{sim}=0.1$ and calculated SS in
separate cells of phase space. These $SS$ vary strong in the
$(p_\|,p_\bot)$ plane
\begin{figure}[thb]
\centerline{\includegraphics[height=5cm,width=7cm,bb=0 0 32mm 25mm]
{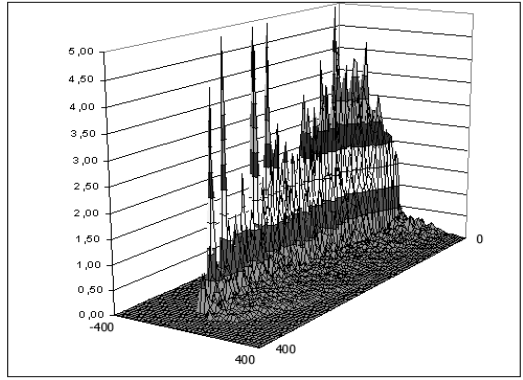}}
\caption{\it A map of $SS$ in $p_\bot$, $p_{||}$ plane for
 $ \sqrt{s}=800 GeV, \lambda_\gamma = -1$, $\Delta k = 0.1$, $\lambda=0$.}
\end{figure}
(example on Figure 1).

The best estimates can be obtained using not the entire phase space but
a region limited with suitable cuts. To find {\it natural cuts}, i.e.
regions of phase space providing the maximal $SS$ value (examples
are shown at Figure 2), the following iterative procedure was used.

On each step the current region is modified by the following rule,
starting with empty region:

\begin{Enumerate}
\item On each step we choose a random phase space cell
(no matter belonging or not to the region found up to the moment)
\item $SS$ value is recalculated for the area with this cell
included (or excluded, if it was already included to the area)
\item If $SS$ increases, this area change is accepted
\end{Enumerate}
This process converges relatively fast (thousands of steps for our
lattice) and results of this procedure are independent
on the choice of the starting cell. The obtained areas are different
for $\Delta k$ and $\lambda$ and depend on energy and photon helicity.

\begin{figure}[thb]
{\footnotesize
    \center{
        \parbox[t]{0.49\textwidth}{\centering{Photon $\lambda_\gamma=-1$}}
        \hfill
        \parbox[t]{0.49\textwidth}{\centering{Photon $\lambda_\gamma=+1$}}
        \\
        \includegraphics[height=5cm,width=7cm]{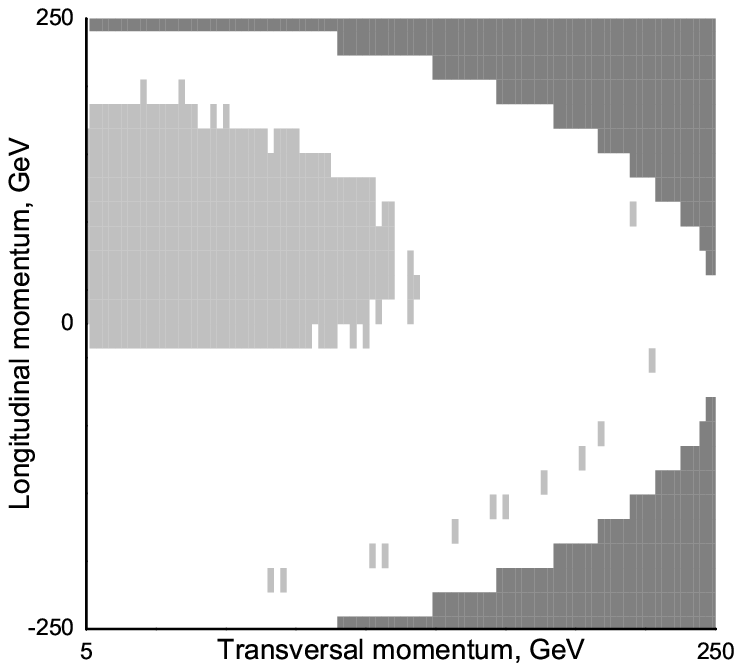}
        \hfill
        \includegraphics[height=5cm,width=7cm]{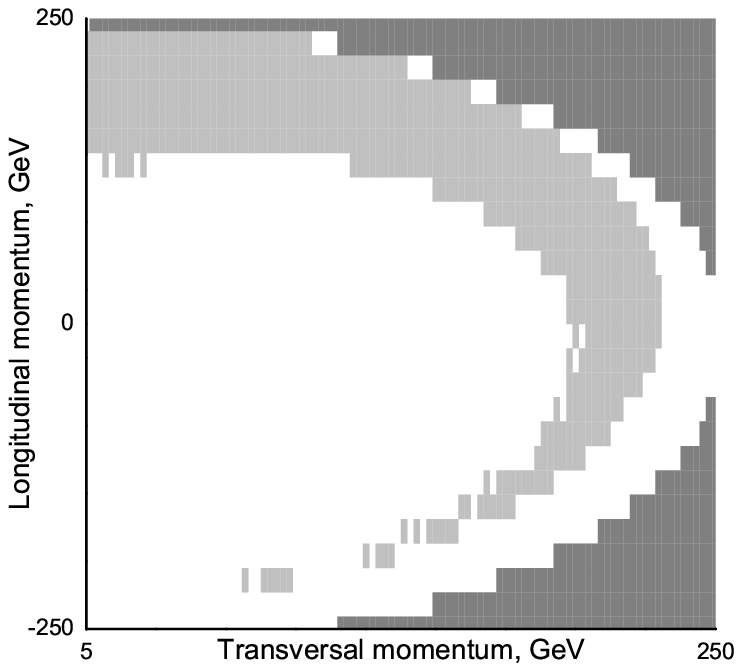}
        \\[-2mm]
        \parbox[t]{0.9\textwidth}{\centering{$\sqrt{s}_{ee}=500$~GeV,
        $\lambda=\lambda_{sim}, \Delta k=0$}\\
        \caption{\it Phase space areas (in muon $p_\bot$, $p_{||}$ plane),
            bringing the best $SS$ value, -- in grey. Dark are
            kinematically forbidden regions.}
 }}}
\end{figure}

The areas responsible for $\lambda$ detection belong to a
small phase space region. One of the essential features is
that reduction of the region to the borders of the phase
space reduces SS slightly (by 10-20\% with momentum cut at 0.7$p_{max}$ ).
With this reduction the intersection of areas,
responsible for two considered anomalies, becomes small. That
means, $\Delta k$ and $\lambda$ can be measured practically
independently.

{\em At the second step} we obtain the final values of the
anomalous parameters achieved in the process by linear
extrapolation with signal level SL (values of signal measured
in $\sqrt{N_{SM}}$ units, fixed by convention) from the
equations\\ \centerline{$\lambda_{exp}=\lambda_{sim}
(SL/SS)$,...}

In the final estimates we also take into account contribution of $e$-channel
as well. For this channel, some new background processes
should be added not present in the $\mu$-- channel. However, their
effect is estimated as small in the areas responsible for
anomalies. Therefore, in preliminary estimates we can account
both $e$ and $\mu$ channels by doubling the number of events
found for $\mu$ channel.

In the Table 2 we compare the results for $e^+e^-\to W^+ W^-$
in all possible channels \cite{TESLA} with muon and electron
channels of $e\gamma\to W\nu$ process.
For this comparison we use $SL = 1$, as used in \cite{TESLA}.
Left column represents here the C.M.S. energy for the initial
$ee$ system.
    \begin{table}[hbt]
\begin{center}
    \begin{tabular}{|@{}p{2cm}@{}|@{}l@{}|@{}c@{}|c|c|} \hline
        $\sqrt{s}_{ee}$, GeV & & $\int\mathcal{L}dt$,$fb^{-1}$
        & $\lambda$ & $\Delta k$ \\
        \hline
        130 & $\mathbf{e\gamma}$ & 100
        & $3.3\cdot 10^{-2}$ & $1.2\cdot 10^{-3}$ \\
        \hline
        500 &  $\mathbf{e\gamma}$
        & 125 & $2.5\cdot 10^{-4}$ & $1.0\cdot 10^{-3}$ \\
        \cline{2-5}
            & $\mathbf{e^+e^-}$& 500
        & $5.9\cdot 10^{-4}$ & $3.3\cdot 10^{-4}$ \\
\hline
        800 &  $\mathbf{e\gamma}$
        & 250 & $1.7\cdot 10^{-4}$ & $1.0\cdot 10^{-3}$ \\
        \cline{2-5}
            &$\mathbf{e^+e^-}$& 1000
        & $3.3\cdot 10^{-4}$ & $1.9\cdot 10^{-4}$ \\
        \hline
    \end{tabular}
\caption{\it Final values of $\Delta k$ and $\lambda$
which can be obtained from $e\gamma\to W\nu$ reaction ($e$ and $\mu$
channels) and from $e^+e^-\to WW$}
\end{center}
\end{table}
Numerical inaccuracy of the shown results is less than 5\%.

These results can be improved by the factor 0.88 accounting
the $\tau$ hadron channel, with branching ratios from \cite{PDG}.

Here the same approach may be used as for simulating $\mu$ channel 2.
Since muon distributions in $\mu$-channels 2 and 1 are similar and
$m_\tau\ll M_W$, one can also use here the data obtained for muon
channels. $\tau$ hadron decay products should have low multiplicity
and small effective mass. Experimentally these events can be
extracted by corresponding cuts ($M_{eff} < m_{\tau}$), and distribution
over total momentum of hadrons should be similar to those for $\mu$ channels
1 and 2. Additional background processes are not likely to appear
under imposed event selection conditions.


Let us summarize our results.
\begin{enumerate}
\item
Even only lepton modes of the process $e \gamma\to\mu\nu$
potentially can provide better opportunities for extracting
$\lambda$, than the process $e^+e^-\to W^+ W^-$; $\Delta k$ is
better measured in the process $e^+e^-\to W W$ (compared to only
lepton modes for $e \gamma\to\mu\nu$). \vspace{-3mm}
\item
In most cases, estimates obtained with unpolarized photons are
irrelevant to the problem. \vspace{-3mm}
\item
The small part of final particles' phase space is responsible
for the best statistical significance for $\lambda$ anomaly.
This part is located near the kinematically allowed border. This
allows one to suppress or even exclude contribution of some
background processes. \vspace{-3mm}
\item
The measured values of $\Delta k$ and $\lambda$ are correlated
weakly.
\end{enumerate}
\vspace{-3mm}

The following steps of the analysis are planned:

1. Due to estimated small influence of backgrounds on the
result, their description can be simplified by approximations
like one used at simulating of the second muon channel
(convolution of production and decay distributions).
Even though the estimates show the backgrounds to influence
the results weakly, these processes will be simulated.
That will lead to more precise predictions and will provide
useful experience for analyzing more complicated processes.

2. $W$-boson quark decay mode should be simulated. We hope
this will also influence the results significantly.\\

We are grateful for helpful discussions to E.Boos, A.Djouadi,
P.Heuer, F.Kapusta, M.Krawczyk, V.Serbo and V.Telnov. This paper
is supported by grants RFBR 99-02-17211, 00-15-96691 and INTAS
00-00679.




\end{document}